\documentclass[12pt,english]{article}
\usepackage{bm,bbm,lscape}
\usepackage[T1]{fontenc}
\usepackage{natbib, amsmath,authblk}
\usepackage{marvosym}
\usepackage{multicol}
\usepackage[british]{babel}
\usepackage{graphicx,xcolor,geometry,setspace,lineno}
\usepackage{pstricks}
\usepackage[british]{babel}
\usepackage[justification=justified,singlelinecheck=false,labelfont=it]{caption}
\usepackage{fancyhdr}
\usepackage{graphicx,epsfig,amssymb,pstricks,amsmath,bm,bbm,natbib,xcolor}

\newcommand{\bfQ}{\hbox{\boldmath$Q$}}
\newcommand{\bfGamma}{\hbox{\boldmath$\Gamma$}}

\newcommand{\bfone}{\hbox{\boldmath$1$}}

\newcommand{\bfpi}{\hbox{\boldmath$\pi$}}

\definecolor{Mygrey}{gray}{0.65}

\title{Semi-Markov Arnason-Schwarz models}

\author{
  Ruth King\footnote{Corresponding author. E-mail: \text{ruth.king@st-andrews.ac.uk}}$\;$ and Roland Langrock\\ \vspace{-0.5em}
  \textit{University of St Andrews}\vspace{-1em}
}

\date{}

\begin{document}

\begin{spacing}{1.1}
\maketitle
%\vspace{4em}
% \begin{linenumbers}
% \setlength\linenumbersep{1cm}
% \rightlinenumbers

%\vspace{-2em}

\begin{abstract}
We consider multi-state capture-recapture-recovery data where observed individuals are recorded in a set of possible discrete states. Traditionally, the Arnason-Schwarz model has been fitted to such data where the state process is modeled as a first-order Markov chain, though second-order models have also been proposed and fitted to data. However, low-order Markov models may not accurately represent the underlying biology. For example, specifying a (time-independent) first-order Markov process assumes that the dwell time in each state  (i.e., the duration of a stay in a given state) has a geometric distribution, and hence that the modal dwell time is one. Specifying time-dependent or higher-order processes provides additional flexibility, but at the expense of a potentially significant number of additional model parameters. We extend the Arnason-Schwarz model by specifying a semi-Markov model for the state process, where the dwell-time distribution is specified more generally, using for example a shifted Poisson or negative binomial distribution. A state expansion technique is applied in order to represent the resulting semi-Markov Arnason-Schwarz model in terms of a simpler and computationally tractable hidden Markov model. Semi-Markov Arnason-Schwarz models come with only a very modest increase in the number of parameters, yet permit a significantly more flexible state process. Model selection can be performed using standard procedures, and in particular via the use of information criteria. The semi-Markov approach allows for important biological inference to be drawn on the underlying state process, for example on the times spent in the different states. The feasibility of the approach is demonstrated in a simulation study, before being applied to real data corresponding to house finches where the states correspond to the presence or absence of conjunctivitis. 
\end{abstract}

\vspace{0.5em}

\section{Introduction}\label{intro}

Capture-recapture studies are often undertaken on wildlife populations. Within these studies, researchers going into the field at a series of capture events. At the initial capture event all individuals observed are uniquely marked, recorded and released. At each subsequent capture event, all individuals observed are recorded, marked if they have not previously been observed and again released. The observation of additional dead individuals leads to capture-recapture-recovery data. Such studies have been formulated within a state-space or hidden Markov model (HMM) framework, e.g., in \citet{gim07}, \citet{roy08} and \citet{king12,king14}, which has proven particularly 
%\comment{The data correspond to the encounter history of each individual observed at least once within the study, detailing whether the given individual is observed or not at each capture event (0 = unobserved; 1 = observed alive; 2/$\dag$ = recovered dead). } 
useful for scenarios which additionally involve individual-level covariate data. We consider the case of time-varying (or dynamic) discrete individual covariates \citep{lebnbps09}. For example, such covariates may relate to breeding status, disease status, foraging/resting, life-stage, mark status, etc. The covariate values are often referred to as `states' and we use these terms interchangeably. Typically it is assumed that if an individual is observed, their corresponding state is also observed. %; however, this need not be the case. % (as we will see in Section \ref{appl}). 
The observed data are the capture/encounter histories for each individual observed within the study. For example, assuming only a single discrete covariate, an observed capture history for an individual may be:
\begin{eqnarray*}
\hspace{1cm} 3 \quad 0 \quad 0 \quad 2 \quad 3 \quad 0 \quad 0 \quad 0 \quad \dag & &
\end{eqnarray*}
This history represents an individual that is observed at the initial capture time in state 3, recaptured at times 4 and 5, in states 2 and 3, respectively, and recovered dead at the final capture occasion. A `0' indicates that an individual is unobserved at the given capture time. 

The Arnason-Schwarz (AS) model is often fitted to such multi-state capture-recapture-recovery data (\citealp{brownie93,schwarz93,kinb03}; for a review see \citealp{lebnbps09}). This model typically assumes the state process to be a first-order Markov chain. Under time-homogeneity, this involves the implicit assumption that the times spent in the states (i.e., the dwell-times) are geometrically distributed, so that the length of time an individual remains in a state is independent of the length of time the individual has already spent in the state (i.e., the process is memoryless). Further, the most likely duration of a stay in a state is one time unit (conditional on an individual being alive). These assumptions facilitate the implementation, but are often not biologically realistic. In particular, the probability an individual leaves a state is often a function of the length of time the individual has been in the state. For example, consider the two states corresponding to `foraging' and `resting' --- an individual will typically forage until its hunger is satiated which in general will be a function of the length of time it has been foraging. Alternatively state may correspond to the disease states `susceptible' and `infected' with the length of time an individual is infected having a non-geometric distribution --- the disease follows a biological process where the recovery time may follow a distribution with mode distinct from one. Higher-order Markov processes can be specified: for example \cite{hestbeck91}, \cite{brownie93} and \cite{roucp09} consider a second-order state process. However, the number of transition parameters increases exponentially as the order increases, so that consideration of processes of order greater than two is usually infeasible and these models only provide limited additional memory structure. Semi-Markov models provide a significantly more flexible yet parsimonious specification, by specifying a general distribution for the state dwell-time distributions. Corresponding models can be interpreted as age-dependent models, where `age' corresponds to length of time spent in the given state. This idea has been applied to (single-state) stopover models (\citealp{pledger09,fewp09}), while \citet{cho13,cho14} considered multi-state capture-recapture models in which {\em one} of the covariate states is assumed to be semi-Markovian.
 
We develop a general class of semi-Markov AS models, where a distribution on the positive integers --- e.g., a Poisson distribution shifted by one (so that the support becomes $1,2,\ldots$), a negative binomial distribution shifted by one, or simply the geometric distribution (in which case the state process is Markovian) --- is specified on the dwell-time for each of the possible states. Together with the conditional state transition probabilities, given an individual being alive and a state being left, this determines the semi-Markov model for the covariate process. Further, we describe an efficient model-fitting approach for the proposed general semi-Markov AS models. The semi-Markov AS model specification is substantially more flexible than a first-order Markov model, yet has only very few (if any) additional parameters. In addition, it is possible to apply standard model selection techniques (e.g., information criteria or likelihood ratio tests) and test for specific biological hypotheses such as the absence or presence of memory in the different states (i.e., comparing first-order memoryless Markov models with semi-Markov memory models).

%Statistical inference in the proposed class of semi-Markov AS models can be done by expanding each semi-Markovian state into a sufficiently large set of Markovian states sharing the same recapture and survival probabilities, and structuring the transition probabilities between those states appropriately. This technique is described in detail in the general context of hidden semi-Markov models (HSMMs) by \citet{lan11}. In particular, the desired dwell-time distributions can be approximated arbitrarily accurately. Thus, it is possible to approximate any semi-Markov AS model with arbitrary accuracy using a first-order Markov AS model. Further, standard hidden Markov model (HMM) techniques can be used to efficiently calculate the likelihood of a semi-Markov AS model, so that model fitting and model selection are straightforward.

The manuscript is structured as follows. %Section \ref{not} provides the mathematical notation used within the manuscript. 
Section \ref{sect2} describes the different model formulations and associated estimation methods. A simulation study is provided in Section \ref{simul}, before we analyse capture-recapture data of house finches, where state corresponds to presence/absence of conjuctivitis in Section \ref{appl}. Section \ref{discuss} concludes with final remarks.

% Multi-state capture-recapture models, i.e., Arnason-Schwarz models, provide a means of estimating both survival rates and probabilities of transitioning between different states, for a given animal population. Such states may correspond to location, but they can also be physiological or behavioural proxies. Usually a first-order Markov process is used to model transitions between different states (and also survival). Under time-homogeneity, this involves the implicit assumption that the times spent in the states are geometrically distributed, and in particular that the most likely duration of a stay in a state is one time unit. Depending on the process to be modelled this clearly can be unrealistic. Here we discuss how to incorporate semi-Markov components in Arnason-Schwarz models, such that the time spent in a state can be modelled using any distribution on the positive integers. This extends the flexibility of the class of Arnason-Schwarz models, and may lead to ecologically meaningful inference on the stochastic structure of the state process. We demonstrate how corresponding models can conveniently be fitted using hidden Markov model techniques. 

\section{Semi-Markov AS model}\label{sect2}

\subsection{Mathematical notation}\label{not}

\subsubsection{Observations}

We let $N$ denote the number of individuals observed within the study, $T$ the total number of capture occasions and $\mathcal{K} = \{1,\dots,K\}$ the set of possible discrete covariate values. For each individual $i=1,\ldots,N$, and capture time $t=1,\ldots,T$, 
\[
x_{it} = \left\{ \begin{array}{cl} 0 & \mbox{if individual $i$ is not observed at time $t$;} \\
k & \mbox{if individual $i$ is observed alive in state $k \in \mathcal{K}$ at time $t$; and} \\
\dag & \mbox{if individual $i$ is recovered dead in the interval $(t-1,t]$.} \end{array} \right. 
\]
%Thus $\bfx_i = \{x_{it}:t=1,\dots,T\}$ corresponds to the capture history of individual $i$, and $\bfx = \{\bfx_i:i=1,\dots,N\}$ the set of all observed capture histories. 
Letting $t_{i0}$ denote the initial capture time for individual $i$, we define $\mathcal{W}_i = \{t \geq t_{i0} : x_{it} \in \{ 1,\ldots,K,\dag \} \}$ as the set of times that individual $i$ is observed, and let $\mathcal{W}_i^c$ denote its complement as the set of times the individual is potentially alive but not observed following their initial capture. We assume that covariate values are observed without error. This assumption will be relaxed in Section \ref{appl:2} where covariates may be unknown when an individual is observed.

\subsubsection{Survival and covariate states}

We now consider the underlying state process and the associated notation. For individual $i=1,\dots,N$, we let $s_{it}$ denote the \emph{true} state of the individual at time $t$, such that, for $t=t_{i0},\dots,T$:
\[
s_{it} = \left\{ \begin{array}{cl}
k & \mbox{ if individual $i$ is alive and has discrete covariate value $k \in \mathcal{K}$ at time $t$;} \\
K+1 & \mbox{ if individual $i$ is dead at time $t$, but was alive at time $t-1$;} \\
K+2 & \mbox{ if individual $i$ is dead at time $t$, and was dead at time $t-1$.} \end{array} \right.
\]
We distinguish ``recently dead'' (state $K+1$) from ``long dead'' (state $K+2$) individuals, since it is typically assumed that individuals can only be recovered dead in the same interval that death occurs. Model adaptations removing this assumption are straightforward and discussed in subsequent sections. %Removal of the assumption is immediate and is discussed with regard to the specification of the observation process (equivalently, this can be achieved by simply considering all individuals in these two categories as a single state of ``dead''). %{\em Roland: I think we need to split state $K+1$ into $K$ different states if we want to allow for site-specific recovery probabilities --- do you agree? Yes} 
For notational convenience we drop the subscript $i$ corresponding to individual, but make it clear in the text that we are referring to the individual. 

\subsection{Traditional AS model}\label{AS1}

\subsubsection{Model parameters}

We initially consider the standard multi-state AS model, which assumes a first-order Markov state process. We allow for both live recaptures and dead recoveries. For $j,k \in \mathcal{K}$, the model parameters are defined as follows:
\begin{eqnarray*}
\phi_t(k) & = & \Pr(s_{t+1} \in \mathcal{K} | s_t = k) \hspace{2.7cm} %= 1 - \mathbb{P}(s_{t+1} = K+1 | s_t = k)  
\mbox{ (\emph{survival probabilities})}; \\
\psi_t(j,k) & = & \Pr(s_{t+1} = k | s_t = j \mbox{ and } s_{t+1} \in \mathcal{K}) \quad \mbox{(\emph{transition probabilities})}; \\
p_{t}(k) & = & \Pr(x_{t} = k | s_{t} = k) \hspace{3.3cm} \mbox{(\emph{recapture probabilities})}; \\
\lambda_{t} & = & \Pr(x_{t} = \dag | s_{t} = K+1) \hspace{2.4cm} \mbox({\emph{recovery probabilities})}; \\
%\alpha_t(k) & = & \mathbb{P}(y_t = k | x_t = 1, s_t = k) \hspace{2cm} \mbox{(\emph{assignment probabilities})}; \\
\pi_{t_0}(k) & = & \Pr(s_{t_0} = k) \hspace{4.4cm} \mbox{(\emph{initial state probabilities})}.
\end{eqnarray*}
%We let $\bfphi = \{\phi_t(k):t=1,\dots,T-1; k \in \mathcal{K}\}$, and similarly for $\bfpsi$, $\bfp$, $\bflambda$, $\bfalpha$ and $\bfpi$. The full set of parameter values is denoted by $\bftheta = \{\bfphi,\bfpsi,\bfp,\bflambda,\bfalpha,\bfpi\}$. 
%The initial state probabilities, %(i.e. $y_{it} = s_{it}$ such that $t \in \mathcal{W}_i$) 
%$\pi_{t_0}(k)$, typically do not need to be estimated; see further comments below. 

\subsubsection{Formulation as HMM}\label{AS1f}

%We initially consider the AS-type multi-state model for capture-recapture-recovery data, specified with a first-order transition process, but allow for the additional complexity that the state may not be recorded when an individual is observed and the possibility that the non recording of the state is state-dependent (i.e. a missing not at random process). 
We extend the notation and HMM-type specification of the Cormack-Jolly-Seber (CJS) model for capture-recapture-recovery data presented by \cite{lank13}. The model is specified as a partially observed state process, coupled with an observation process, defined conditional on the underlying state. %The state and observation processes are a function of the above model parameters. 
The state process can essentially be decomposed into the survival process (i.e., whether or not an individual remains in $\mathcal{K}$) and the transitioning process between covariate states, conditional on survival. The state process has probability mass function (p.m.f.)
\[
f(s_{t+1} | s_t) =  \left\{ \begin{array}{cl} 
\phi_t(s_t) \psi_t(s_t, s_{t+1}) & \mbox{for } s_t, s_{t+1} \in \mathcal{K}; \\
1-\phi_t(s_t) & \mbox{for } s_t \in \mathcal{K},\, s_{t+1} = K+1; \\
1 & \mbox{for } s_t \in \{K+1, K+2\},\, s_{t+1} = K+2; \\
0 & \mbox{otherwise.}
\end{array}
\right.
\]

The p.m.f.\ for the observation process is defined similarly:
%We consider two distinct processes, corresponding to (i) observing the individual and (ii) subsequently recording the associated discrete covariate, conditional on the individual being observed alive. We consider the two processes in turn. 
\[
f(x_t | s_t) = \left\{ \begin{array}{cl}
p_t(s_t) & \mbox{for } x_t = s_t,\, s_t \in \mathcal{K}; \\
1-p_t(s_t) & \mbox{for } x_t = 0,\, s_t \in \mathcal{K}; \\
\lambda_t & \mbox{for } x_t = \dag,\, s_t = K+1; \\
1-\lambda_t & \mbox{for } x_t = 0,\, s_t = K+1; \\
1 & \mbox{for } x_t = 0,\, s_t = K+2; \\
0 & \mbox{otherwise.}
\end{array}
\right.
\]
It is straightforward to modify this definition to deal with scenarios in which ``long dead'' individuals can be recovered \citep[see, for example,][]{catfmn01}, with a time-dependent recovery probability. For example, the recovery probability can be specified as a decreasing function of time following death.
%Note that if ``long dead'' individuals can be recovered \citep[see, for example,][]{catfmn01}, with a time-dependent recovery probability, the only change to the above observation process is to add $f(x_t | s_t) = \lambda_t$ for $x_t = \dag$, $s_t = K+2$, and  $f(x_t | s_t) = 1-\lambda_t$ for $x_{t_0+1},\ldots,x_{t-1} \ne \dag$, $x_t=0$, $s_t = K+2$. (The condition that $x_{t_0+1},\dots,x_{t-1} \ne \dag$ is to ensure that an individual cannot be recovered multiple times). Further recovery models may be specified, such as the recovery probability specified as a decreasing function of time following death.

\subsubsection{Likelihood}\label{1storderlikelihood}

The likelihood is the product over each individual of the corresponding probability of their observed encounter history, conditional on their initial capture. The likelihood component for an individual initially observed at time $t_0$ can be calculated by summing over all possible unknown states after initial capture. Mathematically, 
\begin{eqnarray*}
\mathcal{L} = \sum_{\tau \in \mathcal{W}^c} \sum_{s_{\tau} \in \{1,\dots,K+2\}} \pi(s_{t_0}) \prod_{t=t_0}^{T-1} f(s_{t+1}|s_t)f(x_{t+1}|s_{t+1}). %f(y_{t+1}|s_{t+1},x_{t+1}).
\end{eqnarray*}

This likelihood can be conveniently and computationally efficiently calculated using an HMM-type forward algorithm. In order to see this, we define the time-dependent transition probability matrix (t.p.m.) for the states as
\[
\bfGamma_t(x_t,x_{t+1}) = \left(\begin{array}{ccccc}
\phi_t(1)\tilde{\psi}_t(1,1) & \cdots & \phi_t(1)\tilde{\psi}_t(1,K) & 1-\phi_t(1) & 0 \\
\vdots & & \dots & \vdots & \vdots \\
\phi_t(K) \tilde{\psi}_t(K,1) & \cdots & \phi_t(K) \tilde{\psi}_t(K,K) & 1-\phi_t(K) & 0 \\
0 & & \cdots & 0 & 1 \\
0 & & \cdots & 0 & 1
\end{array}\right)\, ,
\]
where
\begin{equation*}
\tilde{\psi}_t(j,k) = \left\{\begin{array}{cl}
\psi_t(j,k) & \mbox{for } (x_t,x_{t+1}) \in \left\{ (j,k),(0,k),(j,0),(0,0)\right\};\\
0 & \mbox{otherwise.}
\end{array}
\right. 
\end{equation*}
for $j,k \in {\cal K}$. Note that when an individual is observed and the corresponding covariate value recorded at successive times $t$ and $t+1$, the associated $\bfGamma_t$ matrix reduces such that there is only a single non-zero element in the $(K \times K)$ upper-left submatrix. Similarly if the covariate value is observed at time $t$ and not $t+1$, the upper-left submatrix of $\bfGamma_t$ is composed of a single row of non-zero elements; if the covariate value is observed at time $t+1$ and not time $t$ there is a single column of non-zero elements. 

The state-dependent observation process is expressed as a diagonal matrix of the form:
\begin{equation*}
\bfQ_t(x_t) = \left \{ \begin{array}{ll}
\mbox{diag}(1-p_t(1),\dots,1-p_t(K),1-\lambda_t,1) & \mbox{if } x_t = 0; \\
\mbox{diag}(0,\dots,p_t(k),\dots,0,0,0) & \mbox{if } x_t = k \in \mathcal{K}; \\
\mbox{diag}(0,\dots,0,\lambda_t,0) & \mbox{if } x_t = \dag.
\end{array}
\right. 
\end{equation*}

Putting all the pieces together, the corresponding likelihood contribution for the given individual, conditional on being initially observed at time $t=t_0$, is:
\[
\mathcal{L}  = \bfpi_{t_0} \left( \prod_{t={t_0}}^{T-1} \bfGamma_t(x_t,x_{t+1}) \bfQ_t(x_t) \right) \bfone_{K+2},
\]
where $\bfone_{K+2}$ denotes a column vector of ones of length $K+2$, and $\bfpi_{t_0}$ % = (\pi_{t_0}(1),\dots,\pi_{t_0}(K),0,0)$ is a row vector of length $K+2$  and . The row vector 
the initial state distribution at the initial capture, given that the individual is observed. It will often be reasonable to assume that the initial detection does not depend on state, in which case the initial state probability can conveniently be taken as the stationary distribution of the state process conditional on the animal being alive, i.e., the solution to 
\[
\tilde{\bfpi}_{t_0} = \tilde{\bfpi}_{t_0} \left(\begin{array}{ccccc}
{\psi}_t(1,1) & \cdots & {\psi}_t(1,K) \\
\vdots & & \vdots  \\
{\psi}_t(K,1) & \cdots & {\psi}_t(K,K)  
\end{array}\right)\, .
\]
Under this assumption, for an individual observed in state $k \in {\cal K}$ at initial capture, ${\bfpi}_{t_0}$ is the vector of length $K+2$ with $k$-th entry $\tilde{\pi}_{t_0}(k)$ and all other entries 0. Alternatively, the initial state distribution can be estimated. Finally, if we condition on the initial observed covariate value, ${\bfpi}_{t_0}$ is the vector with the $k$th entry equal to one and all other entries 0.

%If we additionally condition on the initial covariate value (as for the standard AS model), or if it is known that initial detections are made only in one particular covariate state, then ${\bfpi}_{t_0}$ is the vector with an entry 1 for the initial covariate value observed and all other entries 0. Finally, ...

% It will often be reasonable to assume that if the covariate value is unobserved at time $t_0$, then 
% ${\bfpi}_{t_0}=(\tilde{\bfpi}_{t_0},0,0)$, where $\tilde{\bfpi}_{t_0}=(\tilde{\pi}_{t_0}(1),\ldots,\tilde{\pi}_{t_0}(K))$ is the stationary distribution of the state process conditional on the animal being alive, i.e., the solution to 
% \[
% \tilde{\bfpi}_{t_0} = \tilde{\bfpi}_{t_0} \left(\begin{array}{ccccc}
% {\psi}_t(1,1) & \cdots & {\psi}_t(1,K) \\
% \vdots & & \vdots  \\
% {\psi}_t(K,1) & \cdots & {\psi}_t(K,K)  
% \end{array}\right)\, .
% \]
% If the covariate is observed to be $j=1,\ldots,K$, at time $t_0$, and under the assumption of stationarity, then ${\bfpi}_{t_0}$ is the vector with $j$-th entry $\tilde{\pi}_{t_0}(j)$ and all other entries 0. Finally, if all initial covariate values are observed and we additionally condition on the initial covariate value (as standard for the traditional AS model) then ${\bfpi}_{t_0}$ is the vector with an entry 1 for the initial covariate value observed and all other entries 0.

%We now extend the multi-state model, removing the first-order Markov assumption for the transition probabilities.

\subsection{Semi-Markov AS model}\label{AS2}

%In particular, let $\delta(r)$ denote the dwell-time for state $r \in \mathcal{K}$. 
%\textbf{Roland - fill in semi-Markov details; initial state distribution (including initial unknown length of time in state) and model-fitting algorithm using approximate first-order Markov model}

\subsubsection{Non-geometric dwell-time distributions}\label{AS2f}

We extend the AS model to allow for semi-Markov state processes, relaxing the restrictive condition that, while alive, the dwell-time distribution in a state follows a geometric distribution (under time-homogeneous transition probabilities). The observation process of a semi-Markov AS model and the survival, recapture and recovery probabilities are defined as for the basic AS model. However, in a semi-Markov AS model, the dwell-time in each state, conditional on survival, can follow any discrete distribution on the natural numbers, e.g., a shifted Poisson or shifted negative binomial. The state process is determined by the survival probabilities, the p.m.f.s of the state dwell-time distributions, $d_1(r), \ldots, d_K(r)$, $r=1,2,\ldots$ --- with associated cumulative distribution functions  $F_1(r), \ldots, F_K(r)$ --- and the conditional state transition probabilities, given the current state is left: 
\[
\psi_t^*(j,k) =  \mathbb{P} \bigl( s_{t+1} = k |  s_t = j \mbox{ and } s_{t+1} \in \mathcal{K}\setminus\{j\} \bigr),
\]
for $j \neq k$. %The difference to the standard AS model is that the probabilities of self-transitions, $\psi_t(i,i)$ in the basic model considered in Section \ref{AS1} 
The probabilities of self-transitions, $\psi_t(k,k)$, conditional on survival, are determined by the dwell-time distributions, 
$d_k(r)$. Consequently, the state process is no longer Markovian, with the probability an individual leaves a state dependent on how long they have been in the state. However, we note that the embedded sequence comprising of simply the order of the states visited --- e.g., the subsequence 1, 2, 1, 3 of the sequence 1, 2, 2, 2, 1, 3, 3 --- does satisfy the Markov property, and is described by the transition probabilities $\psi_t^*(j,k)$. First-order Markov AS models (with time-homogeneous state transition probabilities) are a special case of semi-Markov AS models where all dwell-time distributions are geometric. 
 %We denote the full set of parameter values of the semi-Markov AS model by $\bftheta^* = \ldots $, where ... . {\em (Ruth, as you might recall I'm not a big fan of defining these vectors of parameters --- but happy for you to do it if you are keen. Note this can be a bit tricky if different families of dwell-time distributions are considered in a single model.)}

A common distribution on the positive integers is the shifted negative binomial distribution, which we denote by $\text{sNB}(\nu,\theta)$. The associated p.m.f.\ is given by
$$ p(r) = \begin{pmatrix} r+\nu-2 \\ r-1 \end{pmatrix} \theta^\nu (1-\theta)^{r-1} , \qquad r=1,2,3, \ldots \, .$$
The (shifted) negative binomial distribution represents a generalization of the geometric distribution: $p(r)$ gives the probability that $r-1$ `failures' occur before $\nu$ `successes' have occurred ($\nu=1$ corresponds to the geometric distribution). Using the gamma function, the definition is easily extended for real-valued $\nu$. Other potentially useful dwell-time distributions include the shifted binomial and the shifted Poisson. More flexible specifications, e.g., using mixtures of Poisson distributions, or even such where one parameter is estimated for every point in the support, are equally easy to implement in the framework we propose. % and the shifted quasi-Poisson distribution.

\subsubsection{The likelihood}\label{AS2l}

Inference can be made by extending the approach presented in \citet{lan11} to deal with capture-recapture data, implementing a hierarchical approach of modeling (i) survival state and (ii) covariate states conditional on being alive. More specifically, we formulate an AS model with first-order state process so that it mirrors the properties of the semi-Markov AS model. The corresponding model formulation can be used to fit semi-Markov AS models with {\em any} desired dwell-time distributions. The idea is to use so-called state aggregates, essentially meaning that each of the semi-Markovian covariate states is expanded into a set of Markovian states, structuring the transitions between the latter so that the desired dwell-time distributions are represented. Thus,
we represent an arbitrary semi-Markov AS model (i.e., a hidden semi-Markov model) as a simple AS model (i.e., an HMM) with an expanded state space. The advantage of doing so is that we can use the HMM representation for conducting statistical inference for the given semi-Markov AS model, which means that the entire HMM machinery --- most notably the forward algorithm for evaluating the likelihood --- becomes applicable to the semi-Markov AS model, without any further amendments being required.

Let $a_1,a_2, \ldots , a_K$ be positive integers, and let $a_0=0$. We consider a first-order Markov chain with $a^*=\sum_{l=1}^{K} a_l+2$ states, where state $k \in \mathcal{K}$ of the semi-Markov AS model is expanded into a set of $a_k$ states. We refer to the sets $I_k= \Big\{ n \; \big{\vert} \;  \sum_{l=0}^{k-1} a_l < n \leq \sum_{l=0}^{k} a_l \Big\}$ as {\it state aggregates}, and assume that each state of $I_k$ is associated with the same distribution for the observation process, namely that associated with state $k$ of the semi-Markov AS model. For $k=1,\ldots,K$ and $r=1,2,\ldots$, let $c_k(r)=\{d_k(r)\}/\{1-F_k(r-1)\}$ for $F_k(r-1)<1$, and $c_k(r)=1$ for $F_k(r-1)=1$. The functions $c_k$ play a key role in the following; essentially they are responsible for rendering the desired dwell-time distributions. The $a^* \times a^*$ t.p.m.\ of the model with first-order state process that represents the semi-Markov AS model is defined as:
\begin{equation}
\bfGamma_t^*(x_t,x_{t+1}) = \left(\begin{array}{ccccc}
\phi_t(1) \bfGamma_{11,t}^* & \cdots & \phi_t(1) \bfGamma_{1K,t}^* & (1-\phi_t(1)) \mathbf{1}_{a_1} & 0 \\
\vdots & & \dots & \vdots & \vdots \\
\phi_t(K) \bfGamma_{K1,t}^* & \cdots & \phi_t(K) \bfGamma_{KK,t}^* & (1-\phi_t(K)) \mathbf{1}_{a_K} & 0 \\
0 & & \cdots & 0 & 1 \\
0 & & \cdots & 0 & 1
\end{array}\right) \, . \label{eqnGamma}
\end{equation}
%We now specify $\bfGamma^*_{ij,t}$ for two different cases, corresponding to the diagonal matrices and off-diagonal matrices. 
%\subsubsection*{Case 1: $\bfGamma^*_{ii,t}$, $i=1,\dots,K$}
Here, for $(x_t,x_{t+1}) \in \left\{ (k,k),(0,k),(k,0),(0,0)\right\}$, the $a_k \times a_k$ {\em leading diagonal matrix} $\boldsymbol{\Gamma}_{kk,t}^*$, $k=1,\ldots,K$, is defined as
\begin{equation}\label{bii} \boldsymbol{\Gamma}_{kk,t}^*=
\left(\hspace{-0.5ex} 
\begin{array}{cccccc}
0				& \vrule & {1-c_k(1)}      & 0      &  \ldots  &   0   	 	                          \\
\vdots         		& \vrule & 0                       & \ddots &      	   &   \vdots 	                          \\				           
         		& \vrule & \vdots                  &        &          &   0   	                              \\
0         		& \vrule & 0                       & \ldots &     0    &  {1-c_k(a_k-1)}             \\ 
\hline 
0         		& \vrule & 0                       & \ldots &     0    &  {1-c_k(a_k)}                \\          
\end{array}
\hspace{-0.5ex}\right) \, ;
\end{equation}
else it is the $a_k \times a_k$ matrix with all entries equal to zero. In the special (geometric) case with $a_k = 1$, we define $\boldsymbol{\Gamma}_{kk,t}^*=1-c_k(1)$ for $(x_t,x_{t+1}) \in \left\{ (k,k),(0,k),(k,0),(0,0)\right\}$, and $\boldsymbol{\Gamma}_{kk,t}^*=0$ otherwise. 
%\subsubsection*{Case 2: $\bfGamma^*_{ij,t}$, $i,j=1,\ldots,K$, $i \neq j$}
Furthermore, for $(x_t,x_{t+1}) \in \left\{ (j,k),(0,k),(j,0),(0,0)\right\}$, the $a_j \times a_k$ {\em off-diagonal matrix} $\boldsymbol{\Gamma}_{jk,t}^*$, $j,k=1,\ldots,K$, $j \neq k$, is defined as
\begin{equation}\label{bij} \boldsymbol{\Gamma}_{jk,t}^*=  
\left(\hspace{-0.5ex} 
\begin{array}{cccccc}
\psi_t^*(j,k) c_j(1) & 0 & & \ldots & & 0 \\   
\psi_t^*(j,k) c_j(2) & 0 & & \ldots & & 0 \\   
% &  & &  & &  \\   
\vdots &  & & & &  \\
% &  & &  & &  \\   
\psi_t^*(j,k) c_j(a_j) & 0 & & \ldots & & 0    
\end{array}
\hspace{-0.5ex}\right)\, .
\end{equation} 
In case of $a_k=1$, the zeros disappear. Thus, different dwell-time distributions lead to different $c_k$'s, while the $\psi^*_t(j,k)$'s and the structure of the t.p.m.\ $\bfGamma_t^*(x_t,x_{t+1})$ remain unaffected. %The $c_i$'s are generated solely from the parameters of the dwell-time distributions.
The structure of $\bfGamma_t^*(x_t,x_{t+1})$ can be interpreted as follows. Conditional on survival, all transitions within state aggregate $I_k$ are governed by the diagonal matrix $\boldsymbol{\Gamma}_{kk,t}^*$, which thus determines the dwell-time distribution for $I_k$. The off-diagonal matrices determine the probabilities of transitions between different state aggregates. For example, for $j \ne k$, the matrix $\boldsymbol{\Gamma}_{jk,t}^*$ contains the probabilities of all possible transitions between the state aggregates $I_j$ and $I_k$. Within this model specification these conditional state transition probabilities, given a state is left, are independent of the length of time already spent in a state. % (i.e., independent of the position within the state aggregate). 

We now consider the initial state distribution, $\boldsymbol{\pi}_{t_0}^*$. Using the first-order representation, it is straightforward to fit a semi-Markov AS assuming that the initial capture is not dependent on state and the state process {conditional on the animal being alive}, i.e., the process restricted to the states $\{ 1, \ldots, K\}$, is in equilibrium at the initial capture time, $t_0$. To achieve this, let $\tilde{\bfpi}_{t_0}^*=(\tilde{\pi}_{t_0}^*(1),\ldots,\tilde{\pi}_{t_0}^*(\sum_{l=1}^K a_l ))$ be the solution to the equation system
$$ \tilde{\bfpi}_{t_0}^*  = \tilde{\bfpi}_{t_0}^* \left(\begin{array}{ccccc}
\bfGamma_{11,t}^* & \cdots & \bfGamma_{1K,t}^*  \\
\vdots & & \vdots  \\
\bfGamma_{K1,t}^* & \cdots & \bfGamma_{KK,t}^* 
\end{array}\right) \, .$$
Assuming stationarity of the restricted state process, if the observed covariate is $k=1,\dots,K$ at time $t_0$, then $\boldsymbol{\pi}_{t_0}^*$ is the vector with entries $l \in I_k$ given by $\tilde{\pi}_{t_0}^*(l)$ and all other entries equal to zero. Alternatively, the initial state distribution can be estimated, in which case the equilibrium distribution within state aggregates can be assumed for numerical stability. Finally, if we condition on the initial observed state, we would typically assume stationarity of the aggregated states within the initial state distribution.

%For the case where we condition on the initial covariate value $j=1,\dots,K$, $\boldsymbol{\pi}_{t_0}^*$ corresponds to the vector with entries $i \in I_j$ given by $\frac{\tilde{\pi}_{t_0}(i)}{\sum_{k=1}^{a_i} \tilde{\pi}_{t_0}(k)}$, with all other entries equal to zero. %In other words, this is the vector with non-zero elements corresponding to the normalized elements for the initial state distribution for the aggregates in state $j$. 

%{\em Ruth, I'd like to briefly discuss this before I finish this paragraph.} % by taking the initial distribution $\boldsymbol{\delta}^*$ to be the solution to the linear system $\boldsymbol{\delta}^* \boldsymbol{\Gamma}^* =\boldsymbol{\delta}^*$.  

We now consider the likelihood
\begin{equation}\label{smASlik}
\mathcal{L}  = \boldsymbol{\pi}_{t_0}^* \left( \prod_{t={t_0}}^{T-1} \bfGamma_t^*(x_t,x_{t+1}) \bfQ_t^*(x_t) \right) \bfone_{a^*}.
\end{equation}
The diagonal matrix $\bfQ_t^*(x_t)$ is specified analogously to the definition of $\bfQ_t(x_t)$ in Section \ref{1storderlikelihood}, with diagonal entry $k$, $k=1,\ldots,K$, in $\bfQ_t(x_t)$ repeated $a_k$ times on the diagonal in $\bfQ_t^*(x_t)$, corresponding to the expansion of state $k$ to the state aggregate $I_k$. This representation in general only approximates the semi-Markov model \citep{lan11}. The p.m.f.\ of the distribution of the time spent in state aggregate $I_k$, denoted by $d_k^*(r)$, in general differs from $d_k(r)$ for $r > a_k$, i.e., in the right tail, since by its definition $d_k^*(r)$ exhibits a geometric tail. The difference between $d_k$ and $d_k^*$ can be made arbitrarily small by choosing $a_k$ sufficiently large. For any dwell-time distribution with either finite support or geometric tail, we can ensure that $d_k^*(r)=d_k(r)$ for all $r$ by choosing $a_k$ appropriately. In summary, the likelihood of the semi-Markov AS model can be approximated by (\ref{smASlik}), where the approximation can be made arbitrarily accurate by choosing sufficiently large values $a_k$ for $k=1,\ldots,K$.

\subsection{Inference}

For both the standard AS models and the semi-Markov AS models, for multiple individuals, the likelihood is simply the product of likelihoods of type (\ref{smASlik}), corresponding to each observed capture history. It is then a routine matter to numerically maximize this joint likelihood with respect to the model parameters, subject to well-known technical issues arising in all optimisation problems (e.g., parameter constraints, numerical underflow, local maxima). Approximate confidence intervals for the parameters can be obtained based on the inverse of the estimated information matrix, or, alternatively, using a parametric/non-parametric bootstrap. Model selection, including for the underlying covariate process model, can easily be carried out using model selection criteria such as Akaike Information Criterion (AIC).

\section{Simulation study}\label{simul}

A simulation study was conducted to demonstrate the feasibility of the semi-Markov approach and to investigate potential consequences of fitting simpler Markovian models when the operating model is semi-Markov. We consider a semi-Markov AS model with three underlying covariate states, each with a different dwell-time distribution, namely a shifted negative binomial (with parameters $\nu=4$ and $\theta=0.4$), a shifted Poisson (with mean $\mu=4$) and a geometric (with parameter $\theta=0.4$), respectively. We assume that the other model parameters are constant over time, and so omit the $t$ subscript for each of the parameters. The conditional state transition probabilities (given the covariate state is left) are specified as $\psi^*(1,2)=0.6=1-\psi^*(1,3)$, $\psi^*(2,1)=0.8=1-\psi^*(2,3)$ and $\psi^*(3,1)=0.5=\psi^*(3,2)$. The (state-dependent) survival probabilities are specified as $\phi(1)=0.8$, $\phi(2)=0.9$ and $\phi(3)=0.6$, and recapture probabilities are specified as $p(1)=0.2$, $p(2)=0.1$ and $p(3)=0.5$. The (state-independent) recovery probability is taken to be $\lambda=0.2$.  

\begin{figure}[!tbh]
\begin{center}
{\includegraphics[width=0.9\textwidth]{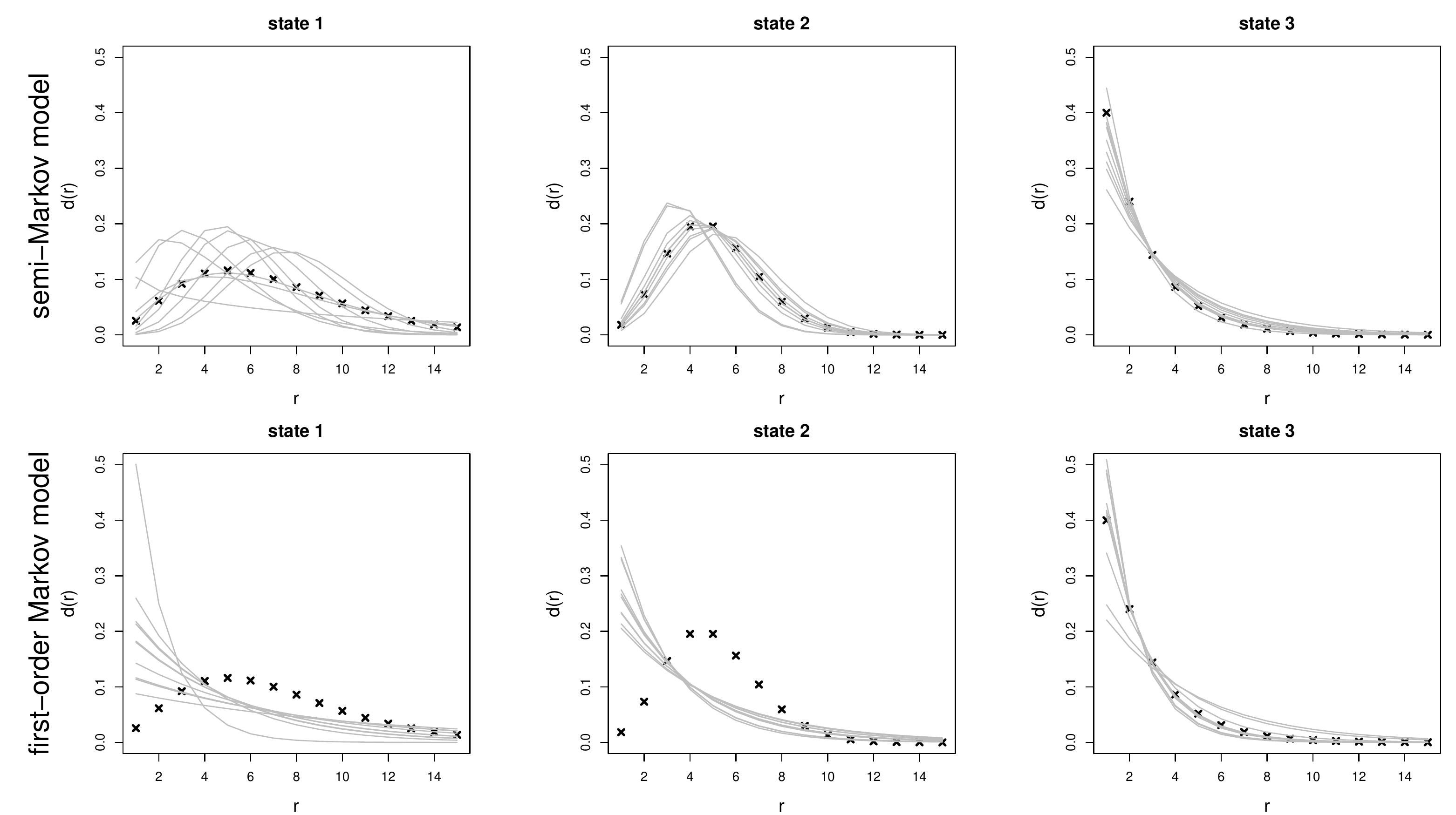}}
\end{center}
\caption{Simulation study: true (black crosses) state dwell-time distributions and the corresponding estimates (grey lines) obtained in the first 10 simulation runs; the top plots show the results obtained when using the semi-Markov model, and the bottom plots those when using the incorrect first-order Markov model. Note that $d(r)$ denotes the p.m.f.\ for the dwell-time distribution, i.e. the probability an individual stays in a given state for $r=1,2,\dots$ capture events. \label{simfig}}
\end{figure}

We simulated 1000 datasets, each with $N=500$ individuals and $T=20$ capture events. For each dataset, we fitted (i) the correctly specified {semi-Markov} AS model and (ii) the corresponding incorrect first-order Markov model. The latter was done in order to investigate potential consequences of not accounting for a semi-Markov structure in the covariate process. Figure \ref{simfig} displays the p.m.f.s of the three dwell-time distributions, together with the estimated distributions obtained in the first 10 simulation runs when using either the first-order or semi-Markov model specification (these are representative of the results obtained for all the simulated datasets). Table \ref{simres} summarizes the results of the remaining parameter estimates obtained using the two different model formulations, by providing the mean relative biases and mean standard deviations.  

\begin{table}[!ht]
\begin{center}
\caption{\label{simres}
MRB: mean relative bias of estimates; MSD: mean standard deviation of estimates. }
\begin{tabular}{ccccccc}
\\
                      & \multicolumn{2}{c}{semi-Markov model} & \multicolumn{2}{c}{first-order Markov model} \\
                      & MRB              & MSD                &  MRB                  & MSD                  \\
\hline
$\hat{\psi}^*(1,2)$   & -0.01            & 0.14               &  0.14                 & 0.15                 \\
$\hat{\psi}^*(2,1)$   &  0.01            & 0.10               &  0.05                 & 0.10                 \\
$\hat{\psi}^*(3,1)$   &  0.01            & 0.22               & -0.16                 & 0.27                 \\
$\hat{\phi}(1)$       &  0.00            & 0.03               &  0.00                 & 0.04                 \\
$\hat{\phi}(2)$       &  0.00            & 0.05               &  0.00                 & 0.05                 \\
$\hat{\phi}(3)$       & -0.01            & 0.08               & -0.01                 & 0.08                 \\
$\hat{p}(1)$          &  0.03            & 0.04               &  0.09                 & 0.07                 \\
$\hat{p}(2)$          &  0.05            & 0.03               &  0.05                 & 0.05                 \\
$\hat{p}(3)$          &  0.11            & 0.19               &  0.15                 & 0.20                 \\
$\hat{\lambda}$       &  0.01            & 0.02               &  0.00                 & 0.02                 \\
\end{tabular}
\end{center}
\end{table}

As expected, the dwell-time distribution is inaccurately estimated when fitting the incorrect geometric distribution (i.e., the first-order Markov model) for the two states with non-geometric distributions. We note that, due to its higher flexibility, the uncertainty in the estimation of the negative binomial state dwell-time distribution (in state 1) was found to be much higher than for the dwell-time distributions in the other two states (cf.\ the top row of plots in Figure \ref{simfig}). In general, the dwell-time distributions of the states may be of interest themselves, e.g., where state refers to life stage which in turn may be used to infer future population trajectories. 

The semi-Markov approach led to approximately unbiased estimates for all the parameters except the recapture probabilities (see Table \ref{simres}). These displayed a positive bias as a result of small-sample properties of the estimation: all mean relative biases decreased to values $<0.005$ when sample size was increased by a factor of 10. We note that the recapture parameters are essentially nuisance parameters, i.e., not of primary interest. In contrast, fitting the first-order model led to biased estimates for the conditional state transition probabilities, despite the fact that the embedded subsequence corresponding to the order 
the states occur in (ignoring duration in each state) is first-order Markovian. Clearly, the inaccurate estimation of the dwell-time distribution and biased estimation of the conditional state probabilities can be problematic when interest lies in the state-switching dynamics. %This may be related to the additional bias observed with regard to the capture probabilities. For example, for the first-order Markov model, the largest relative bias is in the capture probability for state 1, $p(1)$, (which is positively biased), while the conditional transition probability $\psi(3,1)$ is negatively biased. In particular, the over-estimation of the capture probability in state 1 may lead to a reduction in the estimated proportion of individuals moving to state 1 from state 3 to account for the reduced number of birds actually observed in state 1. \textbf{Roland - Does this rationale hold as all of the capture probs are positively biased for the first-order Markov model; but surprisingly the survival probs are not biased - generally the capture and survival probs are negatively correlated....} Roland's response: I would rather not speculate what the reasons for these biases are. Capture probabilities are biased also for the semi-Markov model, where the psi parameters are estimated without bias.
However, interestingly, the survival probabilities were generally robust with regard to the mis-specification of the dwell-time distributions, at least for the set of 
parameter values used in this simulation study.

\section{Application to \emph{Carpodacus mexicanus}}\label{appl}

\subsection{Data}\label{data}

We consider multi-state capture-recapture histories for house finches (\emph{Carpodacus mexicanus}) collected at a study site in Ithaca, New York, between September and December 2002. There are no recoveries in this study. During the study period a total of $N=813$ individuals are observed over the $T=16$ capture occasions. The covariate states correspond to the presence/absence of \emph{Mycoplasma gallisepticum} conjunctivitis, so that $K=2$. We let state 1 denote the absence of conjunctivitis and state 2 the presence of conjunctivitis. For these data, it was not always possible to determine the presence/absence of conjunctivitis (i.e., the state) of an individual bird observed at a given capture occasion: the state is unknown for 4.1\% of the recorded observations. The data are provided in the Web Appendix. Particular interest lies in (i) the survival probabilities for the two states %(it is anticipated that mortality is higher for individuals with conjunctivitis), 
and (ii) the dwell-time distributions for the two states. %, from which we can also estimate the stationary distribution for the proportion of the population with conjunctivitis. 
For further details of the data collection process and the biological mechanisms, see for example, \cite{fau04} and \cite{conc09}.

\subsection{Semi-Markov AS model with unobserved states}\label{appl:2}

We extend the semi-Markov AS model discussed above to account for the possibility of the covariate state of an individual, here presence/absence of conjunctivitis, to be unknown when an individual is observed. %We initially extend the set of possible values observed within a capture history to include an ``unknown'' state. %For individual $i=1,\ldots,N$ and $t=1,\ldots,T$, we define the additional state: $x_{it} = -1$ if individual $i$ is observed at time $t$ but where the associated covariate value is unknown. 
We define $\alpha_t(k)$ to be the assignment probability that the state of an individual is recorded, given that the individual is observed at time $t$ and is in state $k$ (i.e., $s_t = k$). For the special case where all covariates are always observed (without error), the corresponding p.m.f.\ is trivial (a point mass on the true values, so that $\alpha_t(k) = 1$ for all $k \in \mathcal{K}$). Alternatively, if the covariate values are missing at random, then $\alpha_t(k) = \alpha_t$ for all $k \in \mathcal{K}$, and simply corresponds to the probability that the covariate value is recorded independently of state, given the individual is observed. Finally, if the covariate values are \emph{not} missing at random, then in general $\alpha_t(j) \ne \alpha_t(k)$ for $j \ne k$. The corresponding likelihood in the presence of unobserved states is provided in the Web Appendix for both the standard AS model and generalised semi-Markov AS model. The extension to partially observed covariate values, or covariate values observed with error (i.e., false reporting of covariate values) can be similarly incorporated but is omitted here to avoid additional complexity (see \cite{kinm14} for further discussion of such models). 

% \comment{
% We now extend the semi-Markov AS model to the case where covariate values may be observed in their true state or simply be unobserved when an individual is observed (i.e. allowing for $x_t = -1$ for some $t \in \mathcal{W}_i$). In particular, we consider the case where the probability that an individual is observed in a given state is state-dependent (i.e. covariate values are not missing at random, given an individual is observed). The corresponding probability mass function for the observed covariate process is given by:
% \[
% f(y_t | s_t, x_t) = \left\{ \begin{array}{cl} 
% \alpha_t(s_t) & \mbox{for } y_t = s_t \in \mathcal{K},\, x_t = 1; \\
% 1 - \alpha_t(s_t) & \mbox{for } y_t = 0,\, s_t \in \mathcal{K},\, x_t = 1;\\
% 1 & \mbox{for } y_t = 0,\, x_t \in \{0,2\}; \\
% 0 & \mbox{otherwise.} \end{array} \right.
% \]}

\subsection{Models considered}\label{mod}

A first-order Markov AS model, allowing for unknown state when an individual is observed, has been fitted to these data by \cite{conc09}, \cite{schobar11} and \cite{kinm14}. In particular, \cite{kinm14} fit a range of models to the data and identify the following model as optimal via the AIC statistic:
\begin{eqnarray*}
\phi(t+c): & & \mbox{time $(t)$ and covariate $(c)$ dependent survival prob., additive on the logit scale;} \\
p(t+c): & & \mbox{time $(t)$ and covariate $(c)$ dependent capture prob., additive on the logit scale;} \\
\psi: & & \mbox{first-order Markov transition probabilities (time independent);} \\
\alpha(c): & & \mbox{covariate ($c$) dependent assignment probabilities}.
\end{eqnarray*}

We extend the above model to allow for a semi-Markov state process. In particular, we consider each covariate dwell-time distribution to be either a (shifted) negative binomial or a geometric distribution. Thus, we initially consider four possible models. Specifying a (shifted) Poisson distribution led to significantly worse fitting models, so that we omit these results. Since it is not clear if it is reasonable to assume that the covariate process is in its stationary distribution at the initial capture occasion, we
  %condition on the observed initial covariate values when these are known (which is the case for 97.5\% of the encounter histories), and 
tried two different model formulations: 
  %for dealing with those encounter histories for which the initial states are unknown: 
    (i) using the stationary distribution as initial distribution; (ii) estimating an additional parameter giving the probability of being in the conjunctivitis state at the initial capture, but assuming stationarity {\em within} state aggregates. The resulting parameter estimates obtained using the two different models were very close to each other. In particular, for the configuration of dwell-time distributions favored by the AIC, the probability of  being in the conjunctivitis state at the initial capture was estimated as $0.07$ when using (i) and as $0.05$ when using (ii). In the following, we present the results obtained using the simpler model formulation (i).

\subsection{Results}

\begin{figure}[!htb]
\begin{center}
{\includegraphics[width=0.86\textwidth]{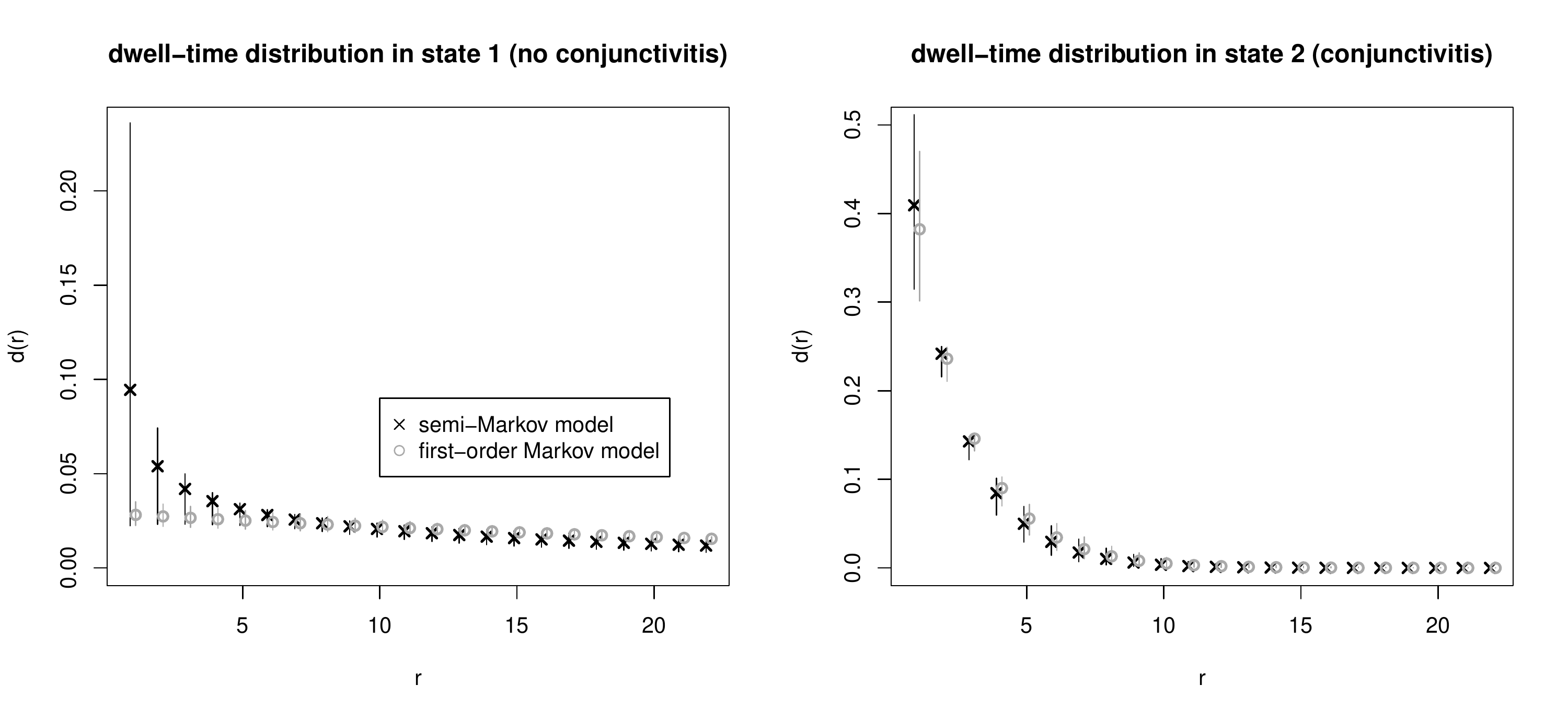}}
\end{center}
\vspace{-1.5em}
\caption{House finch case study: Fitted state dwell-time distributions as obtained using the semi-Markov model (with shifted negative binomial state dwell-time distribution for state 1 and a geometric state dwell-time distribution for state 2; crosses) and the first-order Markov model (circles), respectively. The vertical lines indicate 95\% pointwise confidence intervals obtained using the observed Fisher information. Note that $d(r)$ denotes the p.m.f.\ for the dwell-time distribution, i.e., the probability an individual stays in a given state for $r=1,2,\dots$ capture events. 
\label{dtd}}
\end{figure}

Table \ref{tab1} provides the estimates of the parameters of the dwell-time distributions and the corresponding AIC values for each of the models considered. 
The model deemed optimal specifies a shifted negative binomial dwell-time distribution for state 1 and a geometric dwell-time distribution for state 2 (though the $\Delta$AIC values are small for each of the other models). The fitted dwell-time distributions, for the selected hybrid Markov/semi-Markov AS model and the standard (Markov/Markov) AS model, are displayed in Figure \ref{dtd}. Further, we removed the dependence of the survival and recapture probabilities on time and/or state. Based on the AIC the same structural dependence is retained, though the model with time-constant survival probabilities performed only slightly worse ($\Delta$AIC$=$2.941). %{\em Time-constant $p$s: $5426.438$; time-constant $\phi$s: $5337.673$ (close!); both time-constant: $5490.115$.} 

\begin{table}[!ht]
\begin{center}
\caption{\label{tab1}
Model selection for different dwell-time distributions for the house finch data using the AIC statistic, where state 1 corresponds to the absence of conjunctivitis and state 2 to the presence of conjunctivitis. The MLEs of the parameters for the dwell-time distributions are provided in brackets. The remaining model parameters are of the following form: $\phi(t+k)/ p(t+k) / \alpha(k)$.}
\begin{tabular}{lllc}
\\
\multicolumn{2}{l}{Dwell-time distributions} & \\
\multicolumn{1}{l}{State 1} & \multicolumn{1}{l}{State 2} & AIC & $\Delta$AIC\\
\hline
geom(0.028) & geom(0.382) & 5539.211 & 0.626 \\
sNB(0.581,0.017) & geom(0.409)  & {\bf 5538.585} & {\bf 0.0} \\
geom(0.029) & sNB(0.612,0.287) & 5540.508 & 1.923 \\
sNB(0.536,0.016) & sNB(0.463,0.265) & 5539.144 & 0.559
\end{tabular}
\end{center}
\end{table}

\begin{figure}[!htb]
\begin{center}
{\includegraphics[width=0.9\textwidth]{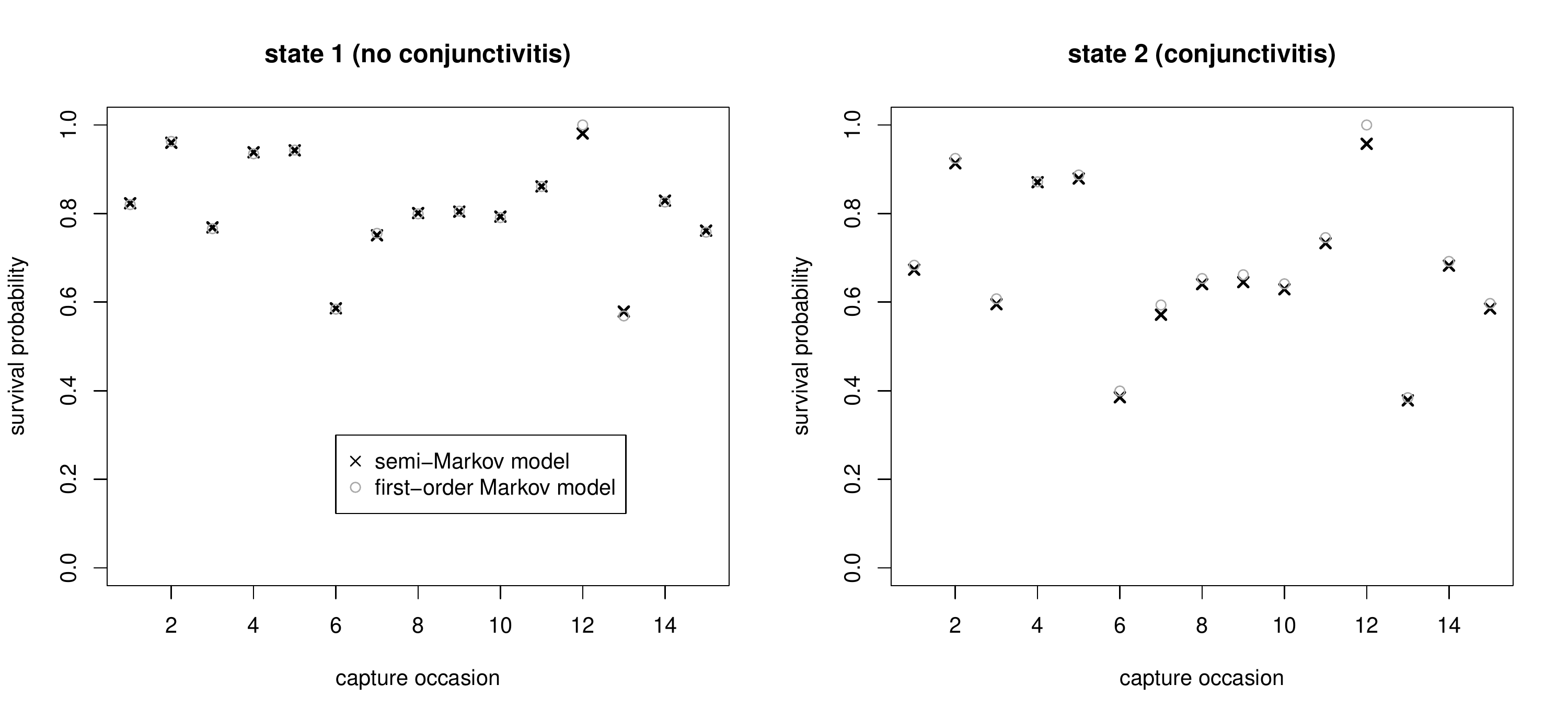}}
\end{center}
\vspace{-1.5em}
\caption{House finch case study: State- and time-dependent survival probabilities estimated using the semi-Markov model (crosses) and the first-order Markov model (circles), respectively. \label{sur}}
\end{figure}

As previously found, the survival probabilities for individuals with conjunctivitis are lower than those for individuals without conjunctivitis present. The differences in the survival probability estimates obtained with the semi-Markov model and the first-order Markov model, respectively, are small (see Figure \ref{sur}). The semi-Markov model predicts higher probabilities for an individual to contract conjunctivitis for short dwell-times in the covariate state corresponding to the absence of  conjunctivitis. This suggests that an individual that has very recently recovered from conjunctivitis has a higher probability of contracting conjunctivitis than an individual that has not suffered from the disease for a longer duration. % \textbf{Roland - may want to check what I have said - basically the memory-less property does not hold? Does this possibly suggest mis-diagnosis of no conjunctivitis present - no symptoms showing....? What do the 95\% CI look like - do they confirm this - I was only looking at the MLEs - worth obtaining the 95\% CIs and see if they overlap or not - and if so state up to which dwell-times?} 
(An alternative explanation may be that of misclassification, where individuals with conjunctivitis are misdiagnosed as healthy --- we do not model possible misclassifications here). 
For the semi-Markov model, the stationary distribution for the states is $(0.933,0.067)$. In other words, the stationary distribution indicates that about 6.7\% of the population will have conjunctivitis at any given time (assuming the individuals observed are representative of the whole population). Assuming a first-order Markov process for both states led to a stationary distribution with 6.9\% of the population with conjunctivitis.

\section{Discussion}\label{discuss}

We extended the AS model for multi-state capture-recapture-recovery data, by permitting the inclusion of memory within the evolution of the states via the specification of a semi-Markov state process. Within this model, memory is incorporated by specifying a parametric distribution on the length of time an individual remains in each given state (i.e., specifying a dwell-time distribution for each state). Corresponding models are attractive due to the increased level of biological realism they permit with a very limited increase in the number of additional parameters needed, in contrast to many previous attempts of introducing memory by using higher-order Markov models. Ignoring memory present within the state process of the model can lead to biased inference, most notably with regard to the conditional state transition probabilities. Interestingly, in our simulation study, no bias was observed in the survival probabilities when fitting an incorrect first-order Markov model. However, in general the dwell-time distributions may themselves be of interest, e.g., where state refers to life-stage or disease status. In such cases the corresponding stationary distribution for the states may also be of interest for future predictions. Further, it is possible to test biological hypotheses with regard to the absence or presence of memory in the state process. 

In our approach to inference in semi-Markov AS models, we have followed the analogous approach to that suggested by \citet{lan11}. The underlying idea is to structure a first-order Markov process such that it accurately represents a semi-Markov process, with the crucial advantage that the efficient recursions available for first-order models become applicable also to semi-Markov models. Various similar approaches have been discussed in the literature. A good overview of these is given in \citet{jon05}, who argues that the usage of this kind of model is ``almost certainly a much better practical choice for duration modeling than development and implementation of more complex and computationally expensive models with explicit modifications to handle duration probabilities''. While the approach in theory could have computational limitations --- as argued by \citet{cho14} --- those would occur only if for some of the dwell-time distributions the support would be very large, covering several hundred time units (i.e., capture occasions). This is unlikely to occur in capture-recapture scenarios, with usually only relatively short time series of capture occasions. 

The semi-Markov AS model was applied to house-finch data, where an indication of memory was found within the dwell-time distribution of an individual being in the state of non-conjunctivitis. In other words, the length of time for which an individual contracts conjunctivitis depends on when it was last infected (i.e., reinfection of conjunctivitis is more likely immediately following their previous recovery from infection). However, the length of time an individual does suffer from conjunctivitis appears to be well modeled via a first-order Markov model (i.e., recovery from conjunctivitis is independent of the length of time an individual has suffered from conjunctivitis). 

The approach described within this paper can immediately be extended to other forms of capture-recapture models including multi-event models \citep{pra05} and partially observed capture-recapture-recovery models \citep{kinm14}, where states are not directly observed or observed with error. We note that the multi-event model has exactly the form of a standard hidden (semi-)Markov model, where no states are directly observed. In these cases, the transition process remains the same, and only the observation process needs to be modified in terms of the assignment probabilities associated with different events or partial states, conditional on being in each given state. This again emphasizes the advantage of specifying the capture-recapture(-recovery) model as a state-space (or hidden Markov) model, separating the system and observation processes \citep{king14}. For example, being able to calculate the likelihood makes it straightforward to estimate model components nonparametrically, via penalized likelihood \citep{mic15}. The extension to multiple covariate processes (e.g., site and breeding status) is also conceptually straightforward, but the curse of dimensionality renders the model-fitting more computationally intensive.

The semi-Markov AS model described currently assumes that the remaining model parameters, most notably the survival probabilities, are time- and/or state-dependent. A similar semi-Markov (state-independent) model can be proposed for the survival probabilities, essentially specifying a parametric distribution on survival, using the analogous approach as that for the state transition probabilities. However, the extension to include a state-dependent semi-Markov survival model is non-trivial if individuals can move between different states. Alternatively, the inclusion of additional covariate information provides a further modeling complexity of interest within such models. The development and fitting of such models is the focus of current research.

% \comment{
% Other possible discussion items:
% \begin{itemize}
% \item state-specific recovery probabilities
% \item semi-Markov process affected by covariates (anything but clear how to do this...)
% \end{itemize}}

%Web Appendix A, containing sample \texttt{R} code for simulating and fitting an AS model involving semi-Markov states will be available with this paper at the Biometrics website. % \texttt{http:www.tibs.org/biometrics}.

\section*{Acknowledgements}

We would like to thank Evan Cooch and Paul Conn for providing the house finch data analysed in Section 4. We would also like to thank the AE and two anonymous referees for their useful comments.

\section*{Appendix: Likelihood for (semi-Markov) AS model with unobserved states}

Here we extend the semi-Markov AS model discussed in Section \ref{AS2} in order to account for the possibility of the covariate state of an individual --- presence/absence of conjunctivitis in the house finch real data case study --- to be unknown when an individual is observed. We expand the set of possible values observed within a capture history to include an ``unknown'' state. For individual $i=1,\ldots,N$ and $t=1,\ldots,T$, we define the additional state: $x_{it} = -1$ if individual $i$ is observed at time $t$ but where the associated covariate value is unknown. Thus, for each individual $i=1,\dots,N$ and capture time $t=1,\dots,T$, the observed data are given by:
\[
x_{it} = \left\{ \begin{array}{rl} 
-1 & \mbox{if individual $i$ is observed at time $t$ but where the covariate state is unknown,} \\
0 & \mbox{if individual $i$ is not observed at time $t$,} \\
k & \mbox{if individual $i$ is observed alive in state $k \in \mathcal{K}$ at time $t$, and} \\
\dag & \mbox{if individual $i$ is recovered dead in the interval $(t-1,t]$.} \end{array} \right. 
\]
As before, for individual $i=1,\dots,N$, we let $s_{it}$ denote the \emph{true} state of the individual at time $t$, such that, for $t=t_{i0},\dots,T$:
\[
s_{it} = \left\{ \begin{array}{cl}
k & \mbox{ if individual $i$ is alive and has discrete covariate value $k \in \mathcal{K}$ at time $t$,} \\
K+1 & \mbox{ if individual $i$ is dead at time $t$, but was alive at time $t-1$,} \\
K+2 & \mbox{ if individual $i$ is dead at time $t$, and was dead at time $t-1$.} \end{array} \right.
\]

We initially describe the likelihood for the AS model in the presence of live recaptures and dead recoveries where covariates may be unknown before extending to the semi-Markov case. 

\subsection*{Standard AS model}

Recall that the likelihood contribution for an individual, conditional on being initially observed at time $t=t_0$, is given by:
\[
\mathcal{L}  = \bfpi_{t_0} \left( \prod_{t={t_0}}^{T-1} \bfGamma_t(x_t,x_{t+1}) \bfQ_t(x_t) \right) \bfone_{K+2}.
\]
Allowing for unobserved covariate states, the observation process is the diagonal matrix:
\[
\bfQ_t(x_t) = \left \{ \begin{array}{ll}
\mbox{diag}(p_t(1)(1-\alpha_t(1)),\dots,p_t(K)(1-\alpha_t(K)), 0 ,0) & \mbox{if } x_t = -1; \\
\mbox{diag}(1-p_t(1),\dots,1-p_t(K),1-\lambda_t,1) & \mbox{if } x_t = 0; \\
\mbox{diag}(0,\dots,p_t(k)\alpha_t(k),\dots,0,0) & \mbox{if } x_t = k \in \mathcal{K}; \\
\mbox{diag}(0,\dots,0,\lambda_t,0) & \mbox{if } x_t = \dag.
\end{array}
\right.
\]
Recall that $\alpha_t(k)$ is defined to be the assignment probability that the state of an individual is recorded, given that the individual is observed at time $t$ and is in state $k$ (i.e., $s_t = k$). 
The transition probability matrix for the covariate process, $\Gamma_t(x_t,x_{t+1})$, remains unchanged except the definition of $\tilde{\psi}_t(j,k)$ is extended such that,
\[
\tilde{\psi}_t(j,k) = \left\{\begin{array}{cl}
\psi_t(j,k) & \mbox{for } (x_t,x_{t+1}) \in \{ (j,k),(0,k),(j,0),(0,0),(-1,k),(j,-1),\\
            & \qquad        (-1,0),(0,-1),(-1,-1) \};\\
0           & \mbox{otherwise,}
\end{array}
\right.
\]
for $j,k \in {\cal K}$. In other words, the set of possible values of $x_t$ and $x_{t+1}$ are extended for which $\tilde{\psi}_t(j,k) = \psi_t(j,k)$, to take into account the additional possibility of unknown covariate values when an individual is unobserved. 

%Finally, we consider the initial state distribution, ${\bfpi}_{t_0}$. As before, we assume that if the covariate value is unobserved at time $t_0$, then ${\bfpi}_{t_0}=(\tilde{\bfpi}_{t_0},0,0)$, where $\tilde{\bfpi}_{t_0}=(\tilde{\pi}_{t_0}(1),\ldots,\tilde{\pi}_{t_0}(K))$ is the stationary distribution of the state process conditional on the animal being alive. If the covariate is observed to be $j=1,\ldots,K$, at time $t_0$, and under the assumption of stationarity, then ${\bfpi}_{t_0}$ is the vector with $j$-th entry $\tilde{\pi}_{t_0}(j)$ and all other entries equal to zero. 

\subsection*{Semi-Markov AS model}

The extension to the semi-Markov case permitting unknown covariate values when an individual is observed follows in the analogous manner as to that presented in Section 2.3 (where covariates are always known when an individual is observed). In particular, each state is expanded into a set of state aggregates and the likelihood corrected accordingly. 

Recall that for the semi-Markov AS model the likelihood contribution for an individual initially observed at time $t_0$ is given by
\[
\mathcal{L}  = \boldsymbol{\pi}_{t_0}^* \left( \prod_{t={t_0}}^{T-1} \bfGamma_t^*(x_t,x_{t+1}) \bfQ_t^*(x_t) \right) \bfone_{a^*}.
\]
The term $\bfQ_t^*(x_t)$ is specified analogous to the definition for the AS model with unobserved states, but allowing for the state aggregates. In other words, the diagonal entry $k=1,\dots,K$ in $\bfQ_t(x_t)$ is repeated $a_k$ times in the diagonal matrix $\bfQ_t^*(x_t)$. The matrix $\bfGamma_t^*(x_t,x_{t+1})$ is defined as in Equation (1) in the main manuscript, but where the leading diagonal matrix $\bfGamma^*_{kk,t}$ is defined as in Equation (2) for the set of observations $(x_t,x_{t+1}) \in \{ (k,k),(0,k),(k,0),(0,0),(-1,k),(k,-1),(-1,0),(0,-1),(-1,-1)\}$. Similarly, the off-diagonal matrix $\bfGamma^*_{jk,t}$ is defined analogously as in Equation (3). %Finally, for the initial state distribution, assuming stationarity of the restricted state process, $\bfpi^*_{t_0} = (\tilde{\bfpi}^*_{t_0},0,0)$ if the initial covariate value is unobserved; else if the covariate is observed in state $j=,\dots,K$ at time $t_0$, $\bfpi^*_{t_0}$ is the vector with entries $i \in I_j$ equal to $\tilde{\pi}^*_{t_0}(i)$ and all other entries equal to zero. 

\end{spacing}

\end{document}